# SDN-Driven Innovations in MANETs and IoT: A Path to Smarter Networks

Andrea Piroddi * and Riccardo Fonti

Department of Engineering and Computer Science, Faculty of Engineering, University of Bologna, Cesena, Italy
Email: andrea.piroddi@unibo.it (A.P.); riccardo.fonti@studio.unibo.it (R.F.)
*Corresponding author

*Abstract*—Mobile Ad Hoc Networks (MANETs) and Internet of Things (IoT) networks operate in decentralized and dynamic environments, making them ideal for scenarios lacking traditional infrastructure. However, these networks face challenges such as inefficient routing, limited scalability, and security vulnerabilities due to their decentralized nature and resource constraints. This paper explores the integration of Software-Defined Networking (SDN) as a unified solution that leverages its centralized control and network programmability to improve routing, resource management, and security. A mathematical model evaluates the impact of SDN integration on Capital Expenditure (CAPEX), Operational Expenditure (OPEX), and performance metrics. Results demonstrate that SDN-enhanced MANETs and IoT networks offer superior scalability, reduced latency, increased throughput, and lower packet loss, especially in dynamic and large-scale environments. While SDN introduces computational overhead, it significantly enhances routing efficiency, resource optimization, and adaptability. The proposed framework provides a robust and scalable solution, enabling the development of network architectures that efficiently manage growing node densities, dynamic topologies, and high data traffic. This approach ensures resilience, making it well-suited to meet the performance and reliability demands of modern, large-scale applications.

## I. INTRODUCTION

Mobile Ad Hoc Networks (MANETs) and Internet of Things (IoT) networks are dynamically formed, infrastructure-less wireless systems designed to operate in highly flexible environments. These networks are particularly suited for critical applications such as disaster recovery, military operations, smart cities, and temporary event coverage [1]. Despite their versatility, both MANETs and IoT networks face challenges such as inefficient routing, scalability issues, and vulnerabilities to attacks like spoofing and Denial-of-Service (DoS) [2, 3].

Recent advancements in Software-Defined Networking (SDN) have shown significant potential to address these challenges by introducing centralized control, programmability, and dynamic resource allocation [4]. By decoupling the control plane from the data plane, SDN enables more efficient routing, enhanced bandwidth utilization, and improved security. These features are particularly advantageous in dynamic environments like MANETs and IoT networks, where traditional distributed protocols often struggle to maintain performance under changing topologies and resource constraints [5, 6].

Although SDN has been widely studied in fixed and cloud-based networks, its application to MANETs and IoT systems remains an evolving area of research. Emerging studies highlight SDN's ability to improve scalability, energy efficiency, and mobility management in these networks [1, 3]. However, integrating SDN into MANETs and IoT networks introduces new challenges, including managing the overhead of centralized control, ensuring compatibility with resource-constrained devices, and adapting SDN controllers to handle highly dynamic and heterogeneous environments [2, 6].

To overcome these limitations, this paper proposes a unified framework that integrates SDN with MANETs and IoT networks. The framework includes a mathematical model to evaluate the impact of SDN integration on Capital Expenditure (CAPEX), Operational Expenditure (OPEX), and performance metrics such as latency, throughput, and packet loss. Furthermore, it explores strategies to address resource constraints, dynamic topology management, and scalability in hybrid MANET-IoT systems.

The remainder of this paper is structured as follows: Section II discusses the characteristics and challenges of MANETs and IoT networks in detail, Section III introduces the proposed SDN integration framework, followed by a mathematical modeling approach in Section IV, results and analysis are presented in Section V, with conclusions and directions for future research in Section VI.

## II. MANETs AND IoT: CHARACTERISTICS AND CHALLENGES

Mobile Ad Hoc Networks (MANETs) and Internet of Things (IoT) networks are distinctive in their ability to form decentralized wireless systems without relying on fixed infrastructure. This capability makes them particularly well-suited for applications such as military operations, disaster recovery, temporary event coverage, smart cities, and industrial automation [1, 3]. However, their flexibility also presents several challenges, particularly in areas such as network control, scalability, and resource management.

A defining characteristic of both MANETs and IoT networks is their ever-changing topology. In MANETs, nodes continuously move, resulting in dynamic routing paths that are frequently created and broken [2]. Similarly, IoT networks often involve devices entering and leaving the network dynamically, or switching operational states based on context [4]. In MANETs, each node may serve as both a host and a router, forwarding packets for others. This dual role requires routing protocols capable of quickly adapting to changes in topology. Traditional protocols such as AODV (Ad-hoc On-Demand Distance Vector) and DSR (Dynamic Source Routing) address these requirements, but they introduce significant overhead in large-scale, highly dynamic networks [5].

The lack of centralized control in traditional MANETs and many IoT implementations often results in suboptimal routing, higher latency, and packet loss during route recalculations. In MANETs, the high mobility of nodes exacerbates these challenges due to frequent link failures caused by rapid node movement. Similarly, in IoT networks, limited computational resources at edge devices can hinder the performance of routing protocols, especially in dense environments with high traffic volumes [6]. Recent studies suggest that incorporating machine learning into routing protocols can improve adaptability to dynamic topologies, though such approaches are computationally intensive and may not be suitable for all IoT devices [5].

Scalability is another critical challenge shared by MANETs and IoT networks. As the number of nodes and their interactions increase, maintaining accurate routing tables and managing network resources become exponentially more complex [3]. In IoT networks, the addition of thousands or millions of devices exacerbates these issues, often leading to interference, congestion, and resource depletion. Additionally, broadcast storms—where excessive control messages flood the network—can overwhelm resources, significantly degrading performance [6].

To address scalability challenges, solutions such as hierarchical routing, zone-based routing, and clustering have been proposed. However, these methods often introduce additional overhead to manage clusters or zones. Recent advancements, including SDN-enabled MANETs and IoT systems, leverage centralized controllers to dynamically manage resources and mitigate the effects of network expansion [4]. While SDN shows promise in improving scalability, it necessitates careful management of controller-node communication overhead, particularly in IoT networks with heterogeneous devices and protocols.

Both MANETs and IoT devices operate in resource-constrained environments, facing limitations in battery life, processing power, and bandwidth [7]. In MANETs, battery consumption is particularly critical, as nodes often rely on finite power sources that are rapidly depleted by continuous route updates and packet forwarding. Similarly, IoT devices, especially those deployed in remote or hard-to-reach areas, must prioritize energy efficiency to ensure long-term operation [3].

Bandwidth constraints further complicate resource management in both MANETs and IoT networks. Shared wireless links often lead to contention, reducing throughput and increasing latency [2]. Effective bandwidth management through techniques like load balancing and congestion control is essential to maintain performance. Emerging solutions, such as cross-layer optimization and SDN-based resource management, have demonstrated significant potential in dynamically allocating resources based on real-time network demands. For instance, cross-layer optimization approaches integrate information from multiple layers of the network stack to adaptively manage resources, improving reliability in dynamic environments [1, 3]. Similarly, SDN-based resource management frameworks enable centralized control and real-time decision-making, as evidenced by recent studies like Shieh *et al.* [8], which demonstrated enhanced scalability and resource allocation in IoT networks. These approaches collectively contribute to improved network reliability and operational efficiency, particularly in scenarios with fluctuating traffic and resource demands.

Addressing these challenges is essential for enhancing the reliability and performance of MANETs and IoT networks. While traditional solutions, such as energy-efficient routing protocols and clustering, have proven effective, the integration of SDN principles offers a more robust and scalable framework for managing dynamic, resource-constrained networks. SDN is described as a powerful toolkit due to its programmability and centralized control, which provide a flexible framework for managing dynamic, resource-constrained networks. This characteristic enables SDN to adapt effectively to the unique demands of IoT and MANET environments, enhancing network efficiency and resilience. For example, recent studies such as Shieh *et al.* [8] have demonstrated how SDN-based management frameworks can optimize resource allocation in IoT networks, improving scalability and reducing latency. Similarly, in MANETs, SDN's ability to centralize control and dynamically adjust routing protocols has been shown to improve network performance under varying traffic conditions, as demonstrated by Wheeb and Al-Jamali [9]. These case studies highlight SDN's adaptability and showcase its potential to address the specific challenges faced by IoT and MANET networks.

III. SDN IN MANETs: INTEGRATION FRAMEWORK

Software-Defined Networking (SDN) introduces a centralized approach to network control, which contrasts sharply with the distributed nature of traditional MANETs and many IoT networks. By decoupling the control plane from the data plane, SDN enables centralized decision-making and holistic control over the entire network. This architecture allows SDN controllers to manage routing, resource allocation, and security policies for MANETs and IoT networks, resulting in improved performance, scalability, and adaptability [1, 4].

### A. Centralized Routing with SDN

In traditional MANETs, routing decisions are made locally by each node using distributed algorithms such as Ad-hoc On-Demand Distance Vector (AODV) or Optimized Link State Routing (OLSR). These protocols often struggle in highly dynamic environments due to frequent route changes and the overhead of route discovery. Similarly, IoT networks, especially those with heterogeneous devices, face challenges in maintaining efficient routing under resource constraints [2].

SDN introduces a centralized controller that computes and optimizes routes based on a comprehensive view of the network topology. Consider a network with nn nodes, where each node has a connectivity degree $d_i$, $i = 1,2,...,n$. In traditional distributed systems, the routing complexity increases proportionally with $d_i$, as each node independently maintains routes to its neighbors. The average path cost $C_{path}$ in such a system can be modeled as:

$$C_{path} = \frac{1}{n}\sum_{i=1}^{n} C_i \qquad (1)$$

where $C_i$ represents the path cost for node i. With increased node mobility, $C_i$ grows due to the need for frequent route recalculations and the overhead of broadcasting route updates [5].

In an SDN-based system, the controller computes routing paths centrally, minimizing redundant route discovery and reducing the overall path cost. This centralized decision-making can be represented as:

$$C_{SDN-path} = \min_{P \in \mathcal{P}} \sum_{i=1}^{n} w_i \cdot d_i \qquad (2)$$

where $w_i$ is a weight assigned based on the traffic load or priority at node $i$, and PP represents all possible paths. The SDN controller dynamically adjusts routes based on real-time traffic conditions, leading to reduced latency and packet loss, as shown in recent studies on SDN-enabled IoT networks [1, 3].

### B. Resource Integration of SDN-Enabled MANETs with IoT

The convergence of Software-Defined Networking (SDN), Mobile Ad Hoc Networks (MANETs), and the Internet of Things (IoT) opens new opportunities for managing dynamic, resource-constrained networks. IoT devices, characterized by limited computational power, bandwidth, and energy resources, present unique challenges that can be addressed through the centralized control and programmability provided by SDN.

In traditional IoT deployments, each device often operates in a static role, relying on pre-configured settings for communication and resource allocation. Integrating IoT nodes into MANETs amplifies the complexity due to the dynamic topology and mobility of nodes. SDN-enabled MANETs offer a solution by centralizing the control of IoT nodes, enabling:

*Dynamic Resource Allocation*: The SDN controller monitors real-time network conditions, such as bandwidth usage, energy levels, and mobility patterns of IoT nodes. By optimizing resource distribution, the controller ensures balanced network performance and prolongs the operational life of resource-constrained IoT devices.

*Efficient Routing*: IoT nodes in MANETs often generate small, periodic data packets, which can create inefficiencies in traditional routing protocols. The SDN controller, with its global view, dynamically adjusts routing paths to minimize latency and reduce packet loss, especially in scenarios with high mobility or interference.

*Energy Management*: Prolonging the battery life of IoT devices is critical. The SDN controller can implement energy-aware routing strategies, prioritizing paths with nodes that have sufficient energy reserves while avoiding overburdening low-energy devices.

In hybrid MANET-IoT environments, where IoT devices coexist with mobile nodes, the SDN controller can facilitate the following optimizations:

*Traffic Differentiation*: IoT applications often generate diverse traffic types, such as periodic sensor updates, event-driven alerts, or multimedia streams. The SDN controller can implement traffic prioritization policies, ensuring that critical IoT data, such as health monitoring alerts, is transmitted with minimal delay.

*Network Slicing*: By logically partitioning the network into slices, the SDN controller allocates resources based on application requirements. For instance, a low-latency slice can be dedicated to real-time IoT applications like autonomous vehicles, while a high-throughput slice supports multimedia streams.

*Load Balancing*: Hybrid networks often experience uneven traffic distribution due to varying node densities or application demands. The SDN controller balances the load across multiple paths, reducing congestion and ensuring fair resource utilization.

The integration of SDN-enabled MANETs with IoT has broad applicability across various domains:

*Smart Cities*: Managing IoT-based infrastructure, such as smart lighting, traffic control, and public safety systems, in dynamically changing environments.

*Disaster Recovery:* Coordinating IoT devices like environmental sensors and drones in emergency scenarios where rapid deployment and adaptability are essential.

*Industrial IoT (IIoT):* Enhancing the performance of manufacturing systems by integrating mobile robots and IoT sensors into a unified, centrally managed network.

### C. Resource Management in SDN-Enabled Networks

Efficient resource management is critical in resource-constrained environments like MANETs and IoT networks, where devices often face limitations in power, bandwidth, and computational capacity [7]. Traditional MANETs rely on decentralized protocols, where each node independently manages its resources. This approach often leads to inefficient resource utilization and network congestion.

In SDN-enabled networks, the controller can dynamically allocate resources based on the real-time demands of the network. Let $B_{total}$ denote the total available bandwidth, and $P_{total}$ the total available power. The allocation cost $C_{alloc}$ can be modeled as:

$$C_{alloc} = \sum_{i=1}^{n} \left( \frac{B_i}{B_{total}} + \frac{P_i}{P_{total}} \right) \tag{3}$$

where $B_i$ and $P_i$ are the bandwidth and power allocated to node $i$, respectively. The SDN controller aims to minimize $C_{alloc}$, ensuring efficient utilization of resources while avoiding overloading any single node [4].

Additionally, SDN controllers implement load-balancing techniques, distributing traffic across multiple paths to prevent congestion. This approach is particularly beneficial in IoT networks, where resource heterogeneity can lead to bottlenecks if not managed effectively [3].

### D. Security Enhancements with SDN

Security remains a significant challenge in both MANETs and IoT networks due to their decentralized nature and the absence of a trusted central authority. Traditional MANETs and IoT systems are highly vulnerable to a range of attacks, including spoofing, blackhole, and Denial-of-Service (DoS) [7]. In these decentralized networks, nodes independently implement security protocols, which can lead to inconsistencies, resulting in exploitable vulnerabilities and potential security breaches.

The adoption of Software-Defined Networking (SDN) offers a promising solution to these security challenges by providing a centralized framework where the SDN controller can enforce uniform security policies across the network. This centralization allows for more effective coordination of security measures, reducing the inconsistencies inherent in traditional decentralized approaches. With SDN, security risks can be managed dynamically, as the controller monitors network activity, detects anomalies, and applies appropriate countermeasures.

To provide a more comprehensive understanding of how SDN mitigates security risks, we can model the security risk $S_{risk}$ in a network as:

$$S_{risk} = \sum_{i=1}^{n} P(V_i) I(V_i) \tag{4}$$

where $P(V_i)$ is the probability of vulnerability $V_i$ being exploited and $I(V_i)$ is the potential impact of such an exploit. The summation ($\sum_{i=1}^{n}$) aggregates the risks across all identified vulnerabilities (*n*). This aligns with risk analysis methodologies that evaluate the cumulative effect of multiple risks on a system or network. The approach of multiplying probability and impact to calculate risk is widely used in risk management frameworks such as ISO 31000 and NIST 800-30. These frameworks emphasize that risk is a function of the likelihood of an event and its consequences, which aligns with our equation. The approach of multiplying probability and impact to calculate risk is widely used in risk management frameworks such as ISO 31000 and NIST 800-30. These frameworks emphasize that risk is a function of the likelihood of an event and its consequences, which aligns with our equation. By structuring the Eq. (4) as a summation, we capture the total risk level of the system, considering all vulnerabilities. This holistic perspective ensures that the derived risk score $S_{risk}$ reflects the aggregate threat posed by all vulnerabilities in the system. This equation is particularly relevant in contexts like IoT and MANET environments, where multiple vulnerabilities exist, and understanding their cumulative risk is essential for designing effective mitigation strategies. This model highlights the need for continuous assessment of vulnerabilities and the potential consequences of their exploitation.

SDN's centralized control provides several key security benefits. First, the controller can dynamically isolate compromised nodes, preventing them from affecting the rest of the network. This rapid isolation is essential in mitigating attacks such as blackhole and DoS, where malicious nodes attempt to disrupt normal operations. Furthermore, the SDN controller can reroute traffic away from vulnerable areas, minimizing the impact of potential security breaches. This feature is particularly useful in MANETs, where node mobility constantly changes the network topology, complicating traditional security measures.

Moreover, SDN can integrate advanced security techniques, such as anomaly detection and intrusion prevention systems, directly into the network infrastructure. By analyzing traffic patterns and node behaviors in real-time, the SDN controller can identify unusual activity indicative of an ongoing attack. These techniques can be customized to the specific security needs of IoT and MANET environments, ensuring that the network remains resilient to evolving threats.

While the centralized nature of SDN provides clear security advantages, it also introduces new challenges. For instance, the SDN controller itself becomes a potential target for attacks, and its security must be rigorously maintained. Ensuring the integrity of the controller's communication with network devices and preventing unauthorized access are critical aspects of SDN security. Additionally, in highly dynamic and heterogeneous IoT environments, where devices have varying capabilities, implementing security measures that are both effective and scalable is a complex task.

Recent studies have examined the effectiveness of SDN-based security measures in MANETs and IoT networks, demonstrating their potential to improve network resilience against common attacks. For example, research has shown that SDN-based intrusion detection systems can significantly enhance attack detection and response times compared to traditional systems [1]. Additionally, SDN's ability to provide real-time, fine-grained control over traffic flows allows for more precise mitigation of network threats.

To further strengthen the security framework, hybrid approaches combining SDN with other security technologies, such as machine learning-based anomaly detection or blockchain for secure data transmission, have shown promise. These approaches leverage the strengths of both SDN's centralized control and the adaptive capabilities of machine learning or blockchain to enhance security further.

In conclusion, while SDN provides a powerful tool for enhancing security in MANETs and IoT networks, its efficacy depends on the careful design and implementation of security measures. Future studies should explore the integration of SDN with other security frameworks, evaluate the performance of SDN-based security measures in real-world environments, and assess the trade-offs between security and performance in large-scale deployments [1, 3].

### E. Adaptive Updates with OpenFlow

Protocols like OpenFlow are central to SDN frameworks, enabling secure and dynamic communication between the control plane and network devices. In the context of MANETs and IoT, OpenFlow helps reduce the complexity of routing updates by providing efficient mechanisms for real-time path optimization. The time $T_{update}$ required to update routing tables in traditional networks can be expressed as:

$$T_{update} = T_{discovery} + T_{propagation} + T_{reconfig} \quad (5)$$

where $T_{discovery}$ is the time for route discovery, $T_{propagation}$ is the time for update propagation, and $T_{reconfig}$ is the time to reconfigure affected nodes. SDN controllers reduce $T_{update}$ by leveraging their global view of the network, minimizing both discovery and propagation times [7].

By integrating SDN into MANETs and IoT networks, it is possible to address critical challenges such as inefficient routing, resource constraints, and security vulnerabilities. The programmability and centralized control offered by SDN enable networks to adapt dynamically to changing conditions, ensuring more efficient and secure operations in complex and heterogeneous environments.

### F. IoT-Specific Challenges and SDN Solutions

While the integration of SDN with MANETs and IoT offers significant advantages, several IoT-specific challenges, such as energy efficiency, sporadic connections, and protocol compatibility, require focused attention. These challenges are particularly crucial in IoT environments where resource constraints and operational limitations must be addressed to ensure sustainable and efficient network performance.

Energy efficiency is one of the most pressing concerns in IoT networks, as many IoT devices are battery-powered and need to operate for extended periods. In traditional IoT systems, devices often have limited energy reserves, and inefficient communication and resource management can lead to rapid depletion of battery life. SDN addresses this challenge by enabling dynamic, energy-aware routing decisions. The SDN controller can monitor real-time power consumption of IoT devices and prioritize routing paths that minimize energy usage. This dynamic energy management not only ensures longer battery life for devices but also contributes to optimizing the overall network performance. The ability of SDN to adjust traffic loads and reduce unnecessary data transmissions is critical in energy-constrained IoT environments.

In IoT networks, especially in mobile or remote environments, devices often experience sporadic connectivity due to network topology changes, interference, or physical barriers. These disruptions can affect the reliability of communication, leading to packet loss and delays. Traditional IoT systems typically rely on static routing protocols that are ill-suited to handle such variability. SDN, on the other hand, benefits from its centralized control, allowing the controller to continuously monitor network conditions and adapt routing paths to avoid areas of poor connectivity. Through real-time adjustments, SDN ensures that IoT devices maintain optimal connectivity, even in challenging environments. This dynamic routing capability significantly improves the robustness and resilience of IoT networks against sporadic connectivity.

IoT networks often involve a variety of devices using different communication protocols, such as MQTT, Zigbee, and CoAP. These protocols are designed for specific applications and have different requirements in terms of data transmission, reliability, and scalability. The diversity of IoT protocols can create compatibility issues, complicating network integration and management. SDN helps to overcome these challenges by providing a centralized framework that can manage and translate between different protocols. The SDN controller can be programmed to detect the protocol requirements of each device and apply the appropriate communication strategy, ensuring seamless interoperability. For instance, SDN can enable efficient communication between MQTT-based devices and Zigbee-based devices by bridging the protocol gap and optimizing data exchange.

In conclusion, while the integration of SDN with IoT presents substantial opportunities, it is crucial to address IoT-specific challenges such as energy efficiency, sporadic connectivity, and protocol compatibility. By leveraging SDN's centralized control and programmability, these challenges can be mitigated, enabling more efficient, reliable, and scalable IoT networks. Future research should continue to explore the integration of SDN with IoT in diverse application scenarios, focusing on energy-conscious network designs, seamless communication across protocols, and robust connectivity solutions.

## IV. MATHEMATICAL MODELING OF COST AND EFFICIENCY

The economic advantages of integrating SDN with MANETs and IoT networks can be assessed using mathematical models that quantify both Capital Expenditure (CAPEX) and Operational Expenditure (OPEX). In this section, we compare the CAPEX and OPEX models for traditional MANETs, IoT networks, and SDN-enabled MANETs, highlighting the cost benefits of the SDN-based approach in both environments. To validate these models, we refer to case studies, such as the integration of SDN in a military communications network, where SDN improved network scalability, efficiency, and resource management in dynamic and large-scale environments, a challenge common to both MANETs and IoT networks [1, 4].

As illustrated in Fig. 1, incorporating SDN into MANETs and IoT networks significantly reduces both CAPEX and OPEX compared to traditional networks. The adoption of general-purpose hardware and centralized control results in reduced overall costs, particularly as the network size and device count grow. The example from the military communications network demonstrates the practical benefits of SDN in reducing capital and operational expenses, making it a viable solution for large-scale IoT deployments, such as in smart cities or industrial automation [3, 4].

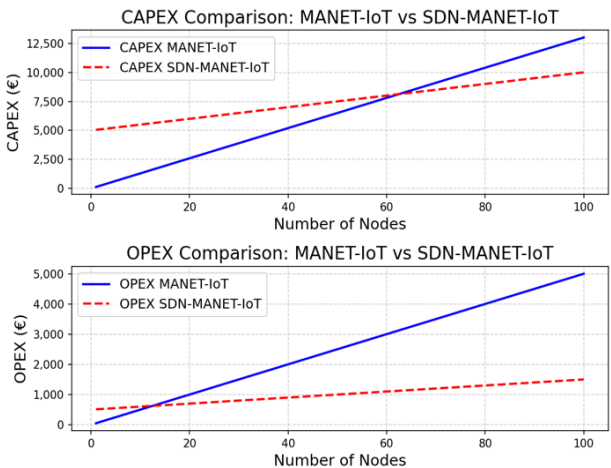

Fig. 1. CAPEX and OPEX efficiency comparison between traditional MANET-IoT and SDN-enabled MANET-IoT frameworks.

### A. CAPEX Model

In traditional MANETs and IoT networks, specialized hardware is required, such as routers, switches, and other network devices capable of independently managing routing, security, and network functions at each node. This decentralized architecture means each node must be equipped with sufficient computational resources and dedicated hardware [7, 9, 10]. Consequently, the CAPEX for traditional networks is highly dependent on the number of nodes $n$ and the cost of hardware components at each node.

For traditional MANETs and IoT networks, the total CAPEX is given by:

$$CAPEX_{MANET/IoT} = \sum_{i=1}^{n} Cost_{hw_i} + \sum_{i=1}^{n} Cost_{sw_i} \quad (6)$$

where $Cost_{hw_i}$ refers to the cost of the hardware at node $i$, including specialized routers, sensors, and other devices, while $Cost_{sw_i}$ refers to the software costs at node $i$, required for routing, security, and management functions [11].

In contrast, SDN-enabled MANETs and IoT networks adopt a centralized control plane. This architecture reduces the reliance on specialized hardware at each node, allowing them to operate with general-purpose hardware, while the SDN controller handles complex computational tasks, including routing and resource management.

The CAPEX for SDN-enabled networks is:

$$CAPEX_{SDN} = \sum_{i=1}^{n} Cost_{hw_i} + Cost_{controller} \quad (7)$$

where $Cost_{controller}$ is the cost of the centralized SDN controller that manages the network. This centralized approach reduces the hardware cost per node by eliminating the need for specialized routing equipment and computational resources at the individual nodes, as demonstrated in large-scale IoT applications like smart grids or environmental monitoring [4].

The primary distinction in CAPEX between traditional MANETs and SDN-enabled MANETs lies in the reliance on general-purpose hardware in SDN-MANETs, which significantly reduces the cost per node. While there is an additional cost for the SDN controller, the overall CAPEX is lower, as the need for specialized hardware at each node is eliminated. For instance, a recent deployment of SDN in a military communications network demonstrated how SDN helped optimize network scalability, leading to lower hardware costs per node, as each node utilized general-purpose computing devices while the SDN controller centralized the network management tasks [4]. As the network size grows, the cost advantages of SDN-MANETs become increasingly apparent, with the per-node hardware cost being lower in SDN-enabled configurations.

### B. OPEX Model

In traditional MANETs and IoT networks, Operational Expenditures (OPEX) are primarily driven by maintenance, configuration, and monitoring costs for each node. Since each node independently handles routing and management tasks, maintaining and updating software and hardware at each node incurs high costs [3, 10]. The total OPEX for traditional networks can be expressed as:

$$OPEX_{MANET/IoT} = \sum_{i=1}^{n}(C_{manutMANET/IoT_i} + C_{monitoringMANET/IoT_i} + C_{configMANET/IoT_i}) \quad (8)$$

where $C_{manutMANET/IoT_i}$ is the maintenance cost at node *i*, $C_{configMANET/IoT_i}$ is the configuration cost, and $C_{monitoringMANET/IoT_i}$ represents the monitoring cost for node *i*. This decentralized approach results in high operational costs, especially in IoT networks with a large number of devices. The Eq. (8) was derived as part of our original analysis to quantify the Operational Expenditure (OPEX) for managing MANET/IoT networks. The derivation process is based on established cost modeling principles and tailored to the unique operational requirements of these networks. The summation captures the total operational expenditure across all *n* nodes in the network. This ensures that the equation accounts for the cumulative cost contributions from every network component, reflecting the distributed nature of MANETs and IoT systems. This equation was independently derived to address the need for a comprehensive OPEX model that considers the unique challenges of dynamic and resource-constrained networks like MANETs and IoT systems. It integrates specific cost factors relevant to these environments, distinguishing it from generic OPEX models in traditional networks. This equation was independently derived to address the need for a comprehensive OPEX model that considers the unique challenges of dynamic and resource-constrained networks like MANETs and IoT systems. It integrates specific cost factors relevant to these environments, distinguishing it from generic OPEX models in traditional networks.

In SDN-enabled MANETs and IoT networks, the centralized controller simplifies network management, reducing the operational costs related to configuration and maintenance of individual nodes. The OPEX for SDN-enabled networks is:

$$OPEX_{SDN} = C_{manutSDN} + C_{configSDN} + C_{monitoringSDN} + \sum_{i=1}^{n} C_{manutNode_i} \quad (9)$$

where $C_{manutSDN}$ is the maintenance cost for the SDN controller, $C_{configSDN}$ is the configuration cost, and $C_{monitoringSDN}$ is the monitoring cost handled by the controller. The term $C_{manutSDN}$ represents the reduced maintenance cost for node *i*, as it no longer needs to perform complex routing or management tasks. This centralized control significantly lowers OPEX, particularly in large-scale IoT applications [7].

### C. Efficiency and Scalability

SDN not only reduces CAPEX and OPEX but also improves the overall efficiency of the network by optimizing routing and resource allocation. In traditional MANETs and IoT networks, the absence of centralized control can lead to suboptimal routing and inefficient bandwidth usage. SDN enables the dynamic and centralized allocation of resources, enhancing network performance. The efficiency of SDN-enabled networks can be expressed as:

$$\eta_{SDN} = \frac{Useful\ Data}{Total\ Bandwidth} \cdot \eta_{optimization} \quad (10)$$

where $\frac{Useful\ Data}{Total\ Bandwidth}$ represents the efficiency of data transmission, and $\eta_{optimization} > 1$ reflects the increase in efficiency due to SDN's ability to optimize traffic dynamically. The Eq. (10) was derived independently to quantify the efficiency of an SDN-based network in terms of bandwidth utilization and the impact of optimization strategies. Here is the rationale behind its formulation. This equation was derived as part of our original analysis to address the need for a specific efficiency metric tailored to SDN environments. Traditional efficiency models often overlook the critical role of SDN's programmability and optimization capabilities. By integrating these aspects into the equation, we aimed to capture the unique contributions of SDN to network performance. In SDN networks, programmability enables dynamic traffic management and resource allocation, directly influencing the effective use of bandwidth. The inclusion of $\eta_{optimization}$ explicitly ties the efficiency metric to the benefits introduced by SDN, making the equation uniquely suited to these networks.

In SDN-enabled MANETs and IoT networks, the centralized control improves bandwidth utilization, reduces delays, and enhances throughput by dynamically adjusting network paths and balancing loads, addressing scalability challenges effectively [3, 4]. The integration of SDN into MANETs and IoT networks leads to substantial reductions in both CAPEX and OPEX, while improving network efficiency, scalability, and security. By centralizing control, SDN enables better resource management and more efficient routing, making it an ideal solution for large-scale, dynamic networks like MANETs and IoT systems, where traditional approaches struggle with the challenges posed by mobility, limited resources, and increasing device density.

In contrast, SDN-enabled MANETs and IoT networks introduce a centralized control plane through the SDN controller, which significantly enhances the network's scalability. The SDN controller has a global view of the network, enabling it to dynamically manage resources, optimize routing paths, and allocate bandwidth in real-time. This reduces the overhead associated with distributed control and allows the network to scale more effectively, especially in large-scale IoT deployments where the number of devices and the complexity of management are key challenges [3, 4].

To model the scalability of SDN-enabled MANETs and IoT networks, we calculated the *network capacity* by considering both the individual node capacities and the additional capacity introduced by the SDN controller. However, this capacity is influenced by the *overhead*, which increases as the number of nodes grows and their mobility leads to frequent topology updates. The model is expressed as:

$$Capacity_{SDN} = \sum_{i=1}^{n} Capacity_i + Capacity_{SDNcontroller} - Overhead \quad (11)$$

where:

- $Capacity_i$ represents the capacity of each individual node $i$, which includes computational power, bandwidth, and routing capabilities.
- $Capacity_{SDNcontroller}$ is the additional capacity provided by the SDN controller, which manages global routing and resource allocation tasks.
- $Overhead$ accounts for the traffic generated by control messages exchanged between the SDN controller and the nodes. This overhead grows with the number of nodes and the complexity of topology updates, which is especially significant in IoT networks with high device density and frequent communication [1, 7].

The overhead was calculated using an algorithm that considers the distances between nodes, the number of nodes, and their mobility. Specifically:

- Each pair of nodes contributes to the number of control packets proportional to their distance and mobility.
- The overhead in bits is computed as:

$$Overhead = Number\ of\ Packets \times Packet\ Size\ (bits) \qquad (12)$$

where the packet size was assumed to be 512 bits (64 bytes). The total overhead is subtracted from the network's capacity, simulating its impact on resource usage and bandwidth. This overhead consideration is particularly relevant in IoT environments, where devices typically have constrained resources, and efficient communication is essential [3].

From the results, it was observed that as the number of nodes increases, the SDN-enabled MANET exhibits higher latency compared to the traditional MANET. This is due to the increased control overhead and the centralization of routing decisions. However, packet loss in SDN-enabled MANET is significantly lower, attributed to the optimized routing paths and better resource management provided by the SDN controller. Furthermore, the throughput in SDN-enabled MANET consistently outperforms that of traditional MANETs, as the centralized control reduces collisions and improves the efficiency of bandwidth usage [4, 12].

Fig. 2 illustrates the results of this model. The graph compares the capacity of traditional MANETs with that of SDN-enabled MANETs, highlighting how the overhead reduces the effective capacity of the latter as the number of nodes increases. Despite this, the improved throughput and reduced packet loss in SDN-enabled MANET demonstrate its potential for more reliable communication in dynamic and large-scale networks, especially in IoT environments where scalability is critical [7].

Fig. 3 illustrates the trend of packet loss for both traditional and SDN-enabled MANETs. The graph highlights the robustness of SDN-enabled MANETs in maintaining low packet loss, even as the number of nodes in the network grows. This demonstrates its suitability for applications requiring high reliability, such as smart cities or industrial IoT systems, where consistent performance is critical despite increasing device density [3, 4].

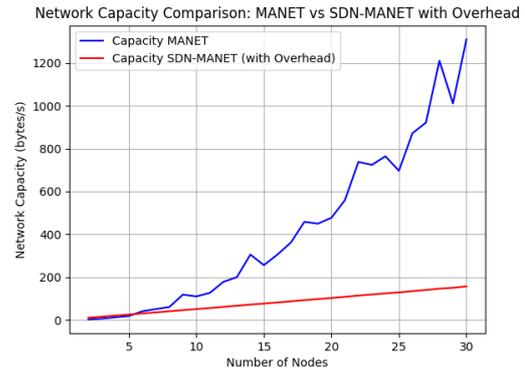

Fig. 2. Comparison of network capacity for MANET and SDN-MANET with increasing number of nodes.

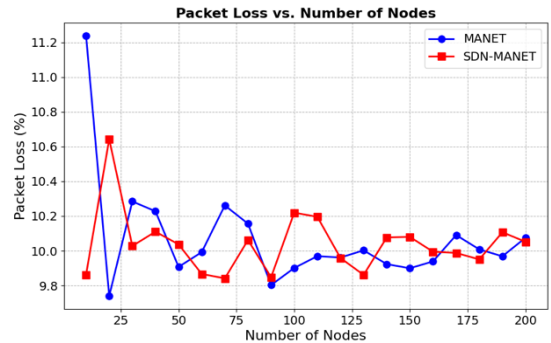

Fig. 3. Trend of packet loss in MANET and SDN-MANET with increasing number of nodes.

Fig. 4 illustrates the latency as the number of nodes increases in both MANET and SDN-MANET configurations.

Fig. 5 illustrates the SDN controller performance as network size increases.

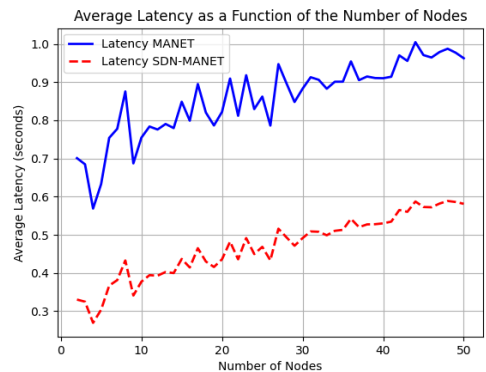

Fig. 4. Latency comparison.

The SDN controller optimizes resource allocation across the network, preventing congestion and ensuring efficient load balancing. This centralized approach is especially effective in mitigating resource constraints as the number of nodes grows, making it highly beneficial for both MANETs and IoT networks, where devices often operate with limited computational power and energy [7, 13]. The controller dynamically computes optimal routes based on a global view of the network, minimizing the number of hops and improving throughput compared to traditional MANETs [4, 12].

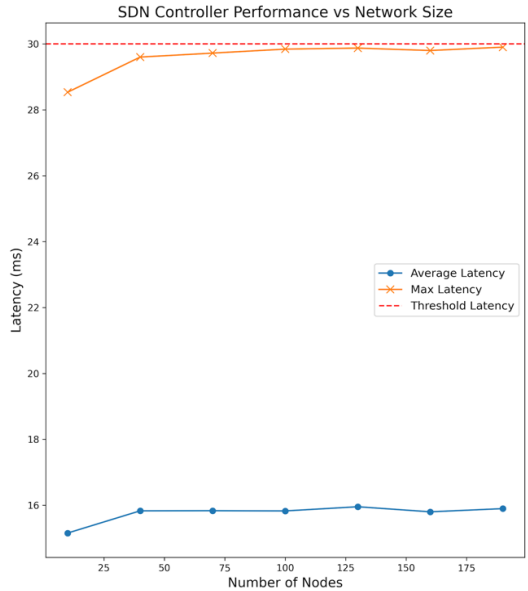

Fig. 5. SDN controller performance *vs* network size.

The SDN controller anticipates topology changes caused by node mobility, recalculating routes and updating flow tables in real-time. This reduces delays and improves responsiveness in dynamic environments, which is particularly important for IoT applications where network conditions can change rapidly due to node mobility or dynamic device behavior [3, 14].

The SDN controller can implement hierarchical or cluster-based management to reduce the complexity of routing and resource allocation in large-scale networks, including IoT systems where the number of devices can grow exponentially. Hierarchical or cluster-based management organizes the network into smaller, manageable groups of nodes, called clusters. Each cluster has a designated leader (or cluster head) that coordinates communication within the cluster and with other cluster heads. This approach reduces the complexity of managing large networks by limiting direct interactions to cluster leaders, which handle inter-cluster communication. This, in turn, improves scalability and resource efficiency, especially in dense or large-scale networks. This approach also enhances scalability by focusing on inter-cluster communication, which is essential in environments like smart cities or industrial IoT networks, where nodes are distributed across different geographical areas and network resources must be allocated dynamically [4, 7].

By enabling logical partitioning of the network, the SDN controller allocates resources dynamically to different types of traffic or services. This slicing improves resource utilization and ensures efficient management of diverse applications such as IoT data streams, security services, and sensor networks. The ability to partition the network logically and assign specific resources based on real-time needs is particularly important for the efficient management of IoT systems, where devices often have different communication requirements [3, 13].

$$Capacity_{total} = Capacity_{clustered} + Capacity_{sliced} \quad (13)$$

The integration of clustering and network slicing further enhances the overall capacity of SDN-enabled MANETs and IoT networks. By abstracting network functions and creating virtual networks, the SDN controller effectively manages larger node counts without overwhelming the network's resources. For IoT, this ensures that millions of devices can be efficiently managed while minimizing interference and congestion in the network [4, 7].

Fig. 6 shows the number of requests waiting to be processed by the SDN controller as the number of nodes in the network increases. As more nodes generate events and submit requests to the controller, the queue size increases, especially if the controller's processing capacity remains fixed. A larger queue indicates that the controller is becoming overwhelmed with requests, potentially leading to delays in processing.

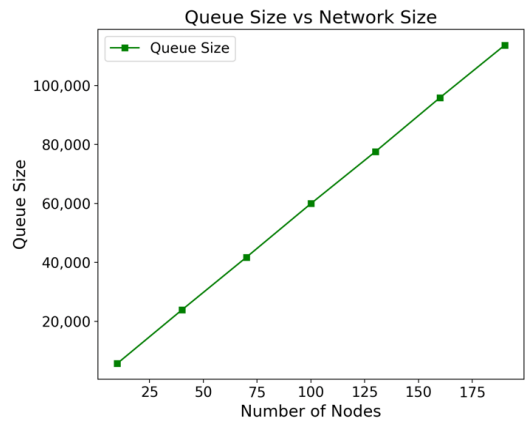

Fig. 6. Queue size *vs* network size.

The security of SDN-enabled MANETs and IoT networks is greatly enhanced by the centralized control provided by the SDN controller. In traditional MANETs, each node is responsible for enforcing its own security policies, which can lead to inconsistencies and vulnerabilities. In contrast, SDN allows for uniform security policies to be enforced across the entire network from a centralized location. The SDN controller can monitor traffic in real-time, detect anomalies, and respond proactively to security threats, which is critical in IoT environments where devices may be subject to cyber-attacks such as spoofing, Man-In-The-Middle (MITM), and Denial-of-Service (DoS) attacks [3, 13]. This centralized monitoring and security enforcement make it easier to detect and mitigate attacks, ensuring that the network remains secure and resilient.

D. *Resource Allocation*

The integration of IoT processes into a Software-Defined Networking (SDN) system necessitates a careful allocation of various resources, including CPU (Central

Processing Unit) memory, network bandwidth, and storage. As the number of IoT nodes increases, the SDN controller must efficiently manage its processing capacity to handle an increasing number of requests. CPU utilization becomes a critical factor, especially when high event rates or complex computations are involved. Memory allocation, similarly, needs to account for the increasing state information for each IoT device, including sensor data, routing tables, and event logs. The growing network traffic due to the influx of IoT devices requires dynamic management of network resources to ensure low latency and prevent congestion, particularly as real-time data streaming and monitoring systems become more prevalent. Storage also plays a key role, as the SDN controller may need to store data temporarily for IoT devices or provide caching and retrieval services for large-scale deployments. Efficient resource allocation strategies that balance these factors are essential for maintaining network performance and scalability in large IoT environments.

Figs. 7–10 provide an insightful view into how the resource utilization of a networked system (CPU, memory, network, and storage) behaves as the number of nodes increases, and they offer valuable implications for understanding network scaling and system performance. Here's a breakdown of the expected results based on the images.

The CPU utilization exhibits a non-linear increase as the number of nodes grows. This is consistent with the idea that an SDN controller processing power is directly impacted by the number of devices it is handling. Initially, as more nodes are added, the CPU utilization rises more slowly. However, due to the non-linear nature of the simulation, the increase in CPU usage accelerates significantly as the number of nodes crosses certain thresholds. This could be interpreted as the SDN Controller needing to process an increasing volume of routing decisions, packet forwarding, and other computational tasks, which leads to a substantial rise in CPU load.

Memory utilization shows a similar non-linear increase with the number of nodes, although its growth rate is slightly slower compared to the CPU utilization. As the network size increases, more memory is required to store routing tables, packet buffers, and network state information. Memory usage generally grows at a moderate pace, as the memory demands are less directly linked to the sheer number of nodes compared to CPU demands. However, as the network becomes larger and more complex, the memory usage gradually increases due to the accumulation of data structures and states.

The network utilization graph also follows a non-linear trend but increases more moderately than CPU and memory utilization. Network utilization typically grows as the number of nodes rises, primarily due to an increase in the volume of traffic and communication demands. While CPU and memory demand increase as each new node adds computational load or data storage requirements, network utilization reflects the additional communication between nodes. The relatively moderate increase in network utilization could suggest that the network is not fully saturated at lower node counts and may reach a point where congestion and traffic bottlenecks will cause more dramatic increases in network load.

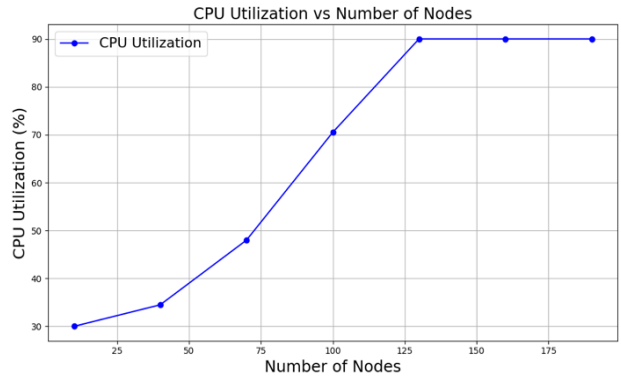

Fig. 7. SDN controller CPU utilization.

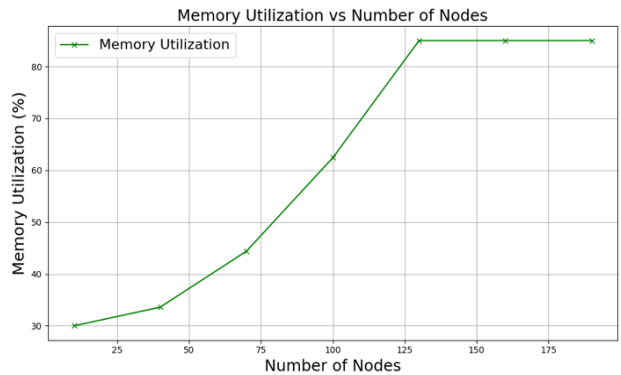

Fig. 8. SDN controller memory utilization.

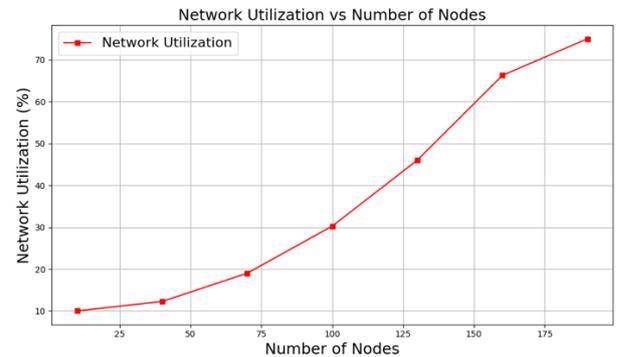

Fig. 9. Network utilization.

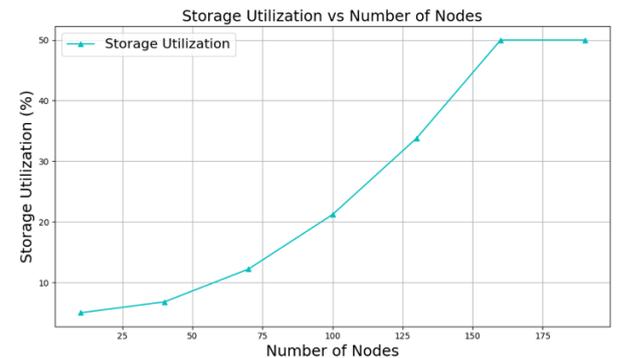

Fig. 10. Storage utilization.

Storage utilization grows slowly with an increase in the number of nodes, as shown in Fig. 9. This behavior is logical, as storage demands are more conservative and often related to data logs, configuration files, and caching rather than the sheer number of devices. As the number of nodes increases, the need for additional storage may rise, but it does so at a much slower rate compared to other resources. This suggests that while network and processing demands scale with the number of nodes, storage requirements often grow at a much slower pace unless specific data-intensive applications are introduced.

These results suggest that as a network grows, CPU and memory become the most critical resources, with their demand scaling non-linearly with the number of nodes. Network and storage utilization increases at more moderate rates but still reflects the growing complexity of the system. Specifically, CPU utilization tends to become the first bottleneck as the number of nodes increases, especially when the SDN Controller needs to handle a significant amount of routing calculations, packet forwarding, and state management for an increasing number of devices. The more efficient the hardware or routing protocols, the less steep the increase in CPU utilization will be.

It is important to note that this simulation does not account for potential issues in the links between nodes, such as packet loss, congestion, or events related to the temporary disappearance and reappearance of network nodes in MANETs. As a result, the curves shown in the graphs do not exhibit fluctuations or oscillations that could arise from such inefficiencies in the links. In a real network environment, such issues could introduce non-linear behaviors and potentially cause peaks or dips in resource utilization curves, particularly in network utilization. Simulations that include factors like link quality and network congestion could therefore present less regular curves, with greater variability in resource utilization values.

In summary, these trends highlight the importance of scalability solutions such as more powerful hardware, load balancing, or specialized hardware accelerators (like ASICs—Application Specific Integrated Circuits) to handle the CPU and memory load as the network grows. Additionally, network design must account for potential bottlenecks due to increased traffic, especially as the network size expands. These observations help design networks that are not only resilient but also efficient in resource allocation as their size and complexity increase.

## V. RESULT AND DISCUSSION

To evaluate the performance benefits of integrating SDN into a MANET, we compare the performance of a traditional MANET with that of an SDN-enabled MANET using key performance metrics such as latency, throughput, Packet Delivery Ratio (PDR), and control overhead. The comparison is based on theoretical models and available simulation results from studies on SDN-MANET integration, including applications to IoT networks where SDN optimizes resource management in dynamic environments [11, 14].

In traditional MANETs, latency is often higher due to the distributed nature of routing and the frequent need for route discovery when nodes move. Every time a route is broken, the route discovery process introduces delays, especially in larger networks. As the network scales and the number of nodes increases, the reactive nature of traditional MANET routing amplifies these delays, making latency a significant challenge. In contrast, SDN-enabled MANETs and IoT networks leverage the centralized SDN controller to maintain a global view of the network and proactively compute optimal routes. This significantly reduces the dependency on reactive route discovery, lowering the overall latency. However, the computational and communication overhead between the SDN controller and the nodes may result in latencies that, while lower, are still comparable to traditional MANETs in certain scenarios. Despite this, SDN routing ensures more consistent and lower latencies, particularly in larger and more dynamic networks [4, 12].

The latency in a traditional MANET can be expressed as:

$$L_{MANET} = L_{route\_discovery} + L_{data\_transmission} \quad (14)$$

where:
- $L_{route\_discovery}$ is the time taken to find a new route when links break or nodes move, and
- $L_{data\_transmission}$ is the time required to transfer data over an established route.

For SDN-enabled MANETs and IoT networks, the latency equation becomes:

$$L_{SDN} = L_{controller\_computation} + L_{data\_transmission} \quad (15)$$

where $L_{controller\_computation}$ computation is the time taken by the SDN controller to recompute routes based on the network's global state, which is generally faster than $L_{route\_discovery}$ in MANETs due to proactive and centralized control. This difference is particularly significant in IoT networks where fast data delivery is essential for real-time applications, such as smart grids or autonomous vehicles [3, 14].

Despite some scenarios where latencies are comparable, SDN-MANETs and SDN-enabled IoT networks generally achieve better latency due to reduced route discovery delays and optimized routing. This improvement is crucial for time-sensitive IoT applications like healthcare monitoring, where low latency is critical to ensure the timely delivery of data from sensors to control systems [4, 7].

Fig. 4 illustrates the observed trends in latency. The graph highlights how SDN-enabled MANETs consistently exhibit lower latency compared to traditional MANETs as the number of nodes increases. This improvement is attributed to the centralized control mechanism of SDN, which significantly reduces delays associated with route discoveries, providing more stable and efficient network performance. While SDN introduces some computational overhead, its proactive routing approach ensures that latencies remain lower and more consistent, even in large and highly dynamic networks, including IoT environments where device density and mobility are high [15, 16].

The analysis in Fig. 4 is based on simulations modeling node mobility and network size, including IoT devices in dynamic scenarios such as smart cities or industrial monitoring systems. In traditional MANETs, latency increases significantly due to frequent route rediscovery. Conversely, in SDN-MANETs, the controller proactively adjusts routing paths, minimizing disruptions caused by node mobility. This advantage is particularly valuable in IoT networks, where devices often operate with limited computational power and require low-latency communication to ensure efficient operation of critical applications such as healthcare monitoring or autonomous systems [17, 18].

- **Impact of Controller Capacity and Network Size on Latency:**

This experiment aims to investigate the relationship between network size, controller capacity, and latency in a Software-Defined Networking (SDN) environment. The primary focus is on understanding how the number of nodes and the limited processing capacity of the SDN controller affect latency, particularly when the network size grows.

The simulation was conducted using a network with 200 nodes, which was progressively scaled by increasing the number of nodes in increments of 30. The network topology was generated using an Erdős–Rényi random graph model with a link probability of 0.05, which creates a sparse graph. This sparsification ensures that the network's path length and complexity increase as the number of nodes grows, thereby making the network more challenging to manage. Each node in the network generated requests at a rate of 20 events per second, and the SDN controller had a processing capacity of 10 requests per second, simulating a scenario where the controller might face a processing bottleneck.

The SDN controller was configured with a maximum acceptable latency threshold of 30 milliseconds, which served as a benchmark for evaluating the system's performance. The simulation ran for a total of 30 seconds to allow the system to reach a steady state and observe the controller's behavior under varying loads. The key metrics collected during the experiment were the average latency, the maximum latency, and the queue size, which reflects the number of pending requests at any given time.

As the number of nodes in the network increased, a clear trend emerged. For smaller networks (up to approximately 40 nodes), the SDN controller could handle the traffic efficiently, keeping both the average and maximum latencies well below the threshold. However, as the network size exceeded 40 nodes, the performance began to degrade. The maximum latency started to approach the 30 ms threshold and eventually reached a point where it asymptotically approached the threshold. This behavior indicated that the controller's processing capacity was becoming overwhelmed by the increasing number of requests, leading to longer processing times and larger queue sizes. Consequently, the system began to experience significant delays in processing requests, as the controller struggled to keep up with the traffic.

This asymptotic approach of maximum latency towards the threshold suggests that a single controller with a processing capacity of 10 requests per second may not be sufficient to handle large-scale SDN networks. When the network size grows beyond 40 nodes, the SDN controller's ability to process requests efficiently becomes constrained, and the latency reaches unacceptable levels. The growing queue size also highlights that the controller is not able to clear requests fast enough, causing a backlog of pending tasks.

These results emphasize the importance of scaling the SDN controller's capacity as the network size increases. In practice, this could mean adding more controllers to distribute the load or enhancing the processing power of the existing controller. Without such scaling, the network will face performance bottlenecks that result in latency violations, as observed in this experiment. To maintain acceptable performance levels, it is crucial to ensure that the controller infrastructure scales with the growing network demand.

In conclusion, see Fig. 5, the experiment illustrates that as the network size increases, the SDN controller's processing capacity becomes a limiting factor. Once the network exceeds 40 nodes, the maximum latency approaches the threshold, indicating that additional controllers or increased processing capacity are necessary to maintain performance. This finding highlights the critical need for scalable SDN controller architectures in large networks.

- **Consideration on Queue Size Growth:**

The queue size in the simulation represents the number of requests that are pending for processing by the SDN controller. It directly reflects the demand on the controller and is influenced by factors such as the event generation rate, the number of nodes, and the controller's capacity to process incoming requests. In this experiment, as the network size increases, so does the queue size, due to the higher number of nodes generating requests at a rate determined by the EVENT_RATE.

As seen in the results, the queue size grows approximately linearly with the number of nodes. This indicates that the demand on the controller increases proportionally as the network expands. For example, when the network consists of 170 nodes, the queue size reaches approximately 100,000, signaling a significant overload in the system. This growth highlights a bottleneck where the controller struggles to handle the increasing number of requests as the network scales.

This behavior suggests that the controller, with its limited processing capacity, may not be sufficient to handle the large volume of requests from a large number of nodes, leading to potential delays and increased latencies. In such cases, introducing additional controllers or increasing the controller's processing capacity could help alleviate the queue size growth and ensure more efficient request handling.

The Packet Delivery Ratio (PDR), defined as the ratio of packets successfully delivered to their destination compared to the number of packets sent, is another key performance metric. In traditional MANETs, frequent route failures and the need for rediscovery often lead to packet losses, negatively impacting the PDR. SDN-enabled MANETs, with their centralized control and ability to adapt quickly to changes in network topology by rerouting packets through alternative paths, generally offer more reliable performance. However, studies indicate that the packet loss rate in SDN-MANETs may remain comparable to that of traditional MANETs under certain conditions, especially in dense IoT deployments where high interference is present [15, 19].

$$PDR_{SDN} \approx PDR_{MANET} \qquad (16)$$

This equivalence suggests that while SDN improves traffic management and route recalculation efficiency, it does not necessarily result in significantly lower packet loss in environments with high node mobility or dense IoT device deployments. Optimizing packet loss in such scenarios requires additional strategies, such as cross-layer optimization or the integration of machine learning techniques for predictive routing [20].

Control overhead refers to the bandwidth consumed by control messages required for network management. In traditional MANETs, a significant portion of the available bandwidth is used for control messages, such as route discovery and maintenance in protocols like AODV or OLSR. This overhead increases significantly in large, dense, or highly mobile networks, including IoT ecosystems where frequent communication between devices amplifies the overhead.

In SDN-enabled MANETs and IoT networks, the centralized controller reduces the number of control messages required for routing decisions, as route computation is handled centrally rather than being distributed across nodes. This leads to lower control overhead:

$$O_{SDN} < O_{MANET} \qquad (17)$$

where $O_{SDN}$ is the control overhead in an SDN-enabled MANET or IoT network, and $O_{MANET}$ is the overhead in a traditional MANET. The reduction in overhead leaves more bandwidth available for data transmission, further improving network performance in IoT applications that require high throughput and low latency, such as video streaming or industrial automation [3, 7, 21].

Table I highlights the performance comparison between traditional MANETs, SDN-enabled MANETs, and IoT networks across key metrics, demonstrating how SDN integration addresses the inherent limitations of these architectures. Traditional MANETs and IoT networks often face challenges such as high latency, low throughput, and limited scalability due to their decentralized nature. IoT networks are further constrained by resource limitations, including energy, bandwidth, and computational capacity.

TABLE I. PERFORMANCE COMPARISON: MANET, SDN-MANET, AND IOT NETWORKS

| Metric | Traditional MANET | SDN-MANET | IoT Networks |
| --- | --- | --- | --- |
| Latency | High, due to distributed routing and frequent route rediscovery [4, 7]. | Low, due to centralized control and proactive routing; may increase slightly with computational overhead [3, 13]. | Low in static deployments; increases in dynamic scenarios without SDN-based optimization [3]. |
| Throughput | Low, as distributed routing may result in congestion and inefficient resource usage [6, 9]. | High, with dynamic load balancing and optimized routing paths managed by the SDN controller [7]. | Moderate; depends on protocol and resource constraints; SDN improves throughput in hybrid MANET-IoT setups [14]. |
| Packet Delivery Ratio (PDR) | Prone to frequent failures due to route breakages and rediscovery delays [7, 11]. | Comparable to traditional MANETs under high interference but improves in dynamic scenarios with adaptive routing [5]. | Often moderate; reliable in small-scale static deployments but decreases with network growth or mobility [17]. |
| Control Overhead | High, as nodes exchange frequent control messages for route discovery and updates [10, 15]. | Low, as the SDN controller handles routing centrally, reducing message exchanges among nodes [4, 11]. | High in dynamic networks due to periodic updates; mitigated by SDN controllers [7, 19]. |
| Scalability | Limited, with network congestion and performance degradation in dense or large-scale deployments [1, 14]. | High, as centralized management ensures efficient resource allocation and routing even in large-scale deployments [4, 12]. | Moderate; often constrained by resource availability and protocol design [3, 16]. |
| Security | Decentralized, prone to attacks such as spoofing, DoS, and unauthorized access [6, 7]. | Enhanced, with uniform security policies and centralized anomaly detection by the SDN controller [3, 4]. | Often basic; requires additional measures for security in dynamic environments [17, 18]. |
| Energy Efficiency | High consumption due to frequent routing and control operations [5, 10]. | Improved, with energy-aware routing and centralized resource optimization [3, 13]. | Limited; battery constraints impact operations; SDN can extend device life in hybrid environments [3, 5]. |

SDN-enabled MANETs introduce a centralized control plane that optimizes routing, resource allocation, and security policies. This results in significantly reduced latency, higher throughput, and improved energy efficiency. In hybrid MANET-IoT environments, where IoT devices coexist with mobile nodes, SDN's centralized management ensures balanced resource distribution, real-time traffic differentiation, and efficient network slicing. These capabilities are critical for applications requiring low-latency communication, such as healthcare monitoring and autonomous vehicles, as well as scenarios demanding high throughput, like Industrial IoT (IIoT) deployments.

The table underscores SDN's ability to reduce control overhead, a frequent bottleneck in traditional MANETs and IoT networks, by centralizing route computation and minimizing inter-node communication. Furthermore, SDN improves scalability by managing large, dense networks with hierarchical clustering and load balancing techniques. Security enhancements offered by SDN, including centralized anomaly detection and uniform policy enforcement, make it particularly valuable in IoT deployments vulnerable to attacks like spoofing and unauthorized access.

By addressing these limitations, SDN integration provides a robust framework for hybrid MANET-IoT networks, ensuring efficient and secure operation even in dynamic and resource-constrained environments. This positions SDN-enabled MANETs as a transformative solution for modern IoT applications, including smart cities, disaster recovery, and industrial automation [4, 12].

IoT and MANETs face critical security challenges due to their decentralized nature, lack of inherent trust mechanisms, and resource constraints. Common vulnerabilities include spoofing attacks, blackhole attacks, Denial-of-Service (DoS) attacks, and data interception. These challenges are exacerbated in hybrid MANET-IoT environments where heterogeneous devices operate in dynamic and resource-constrained conditions.

Common Threats in MANET-IoT Networks:

Spoofing attacks: Malicious nodes impersonate legitimate devices, disrupting communication or capturing sensitive data.

Blackhole attacks: A rogue node misdirects traffic by falsely advertising optimal routes, leading to data loss.

Denial-of-Service (DoS) Attacks: Attackers flood the network with traffic, overwhelming devices and degrading service.

Eavesdropping: Unauthorized nodes intercept data packets, compromising privacy and confidentiality.

SDN as a Security Solution

The centralized control and global visibility provided by SDN address these threats effectively.

Anomaly Detection and Mitigation:

The SDN controller continuously monitors network traffic and uses machine learning or anomaly detection algorithms to identify suspicious patterns indicative of spoofing or DoS attacks.

Once an attack is detected, the controller can isolate the malicious node by rerouting traffic or denying its access.

Dynamic Security Policies:

SDN enables the deployment of uniform security policies across the network. For example, Access Control Lists (ACLs) can be dynamically updated to block malicious IPs or suspicious nodes.

The centralized nature of SDN allows for real-time updates to security rules, ensuring consistent protection across IoT and MANET devices.

Encryption and authentication:

SDN integrates IoT-specific lightweight protocols like Datagram Transport Layer Security (DTLS) and CoAP with enhanced encryption to secure data transmission.

The SDN controller can enforce device authentication mechanisms, such as mutual authentication, to prevent unauthorized access.

Resilience to Blackhole Attacks:

By maintaining a global view of the network, the SDN controller verifies advertised routes before updating routing tables, mitigating the risk of blackhole attacks.

Improved Privacy:

The SDN controller can anonymize sensitive data and ensure secure storage or transmission, reducing the risk of eavesdropping.

Recent studies have demonstrated the effectiveness of SDN in improving network security:

Anomaly detection: SDN controllers equipped with machine learning algorithms achieve detection rates exceeding 95% for DoS and blackhole attacks in dynamic IoT networks [22, 23].

Reduced downtime: By isolating compromised nodes in real time, SDN reduces network downtime by up to 40% in hybrid MANET-IoT deployments [24].

Enhanced data protection: Encryption and authentication mechanisms integrated via SDN result in a 70% reduction in successful spoofing attempts compared to traditional networks [24].

While SDN significantly enhances security, challenges remain:

Controller bottlenecks: The centralized nature of SDN makes the controller a potential single point of failure. Future work should explore redundant and distributed SDN controllers for resilience.

Overhead in resource-constrained devices: IoT devices with limited computational resources may struggle to handle complex encryption or frequent updates from the SDN controller. Lightweight security protocols tailored for SDN-IoT integration are needed.

By addressing these challenges and leveraging SDN's capabilities, hybrid MANET-IoT environments can achieve robust and scalable security frameworks that meet the demands of modern applications.

## VI. CONCLUSION

This study presents a Software-Defined Networking (SDN)-driven framework to address the challenges of Mobile Ad Hoc Networks (MANETs) and Internet of Things (IoT) systems, including scalability, cost efficiency, and resource management. By leveraging SDN's centralized control and programmability, the proposed approach enhances routing efficiency, dynamic resource allocation, and scalability while reducing computational overhead on individual nodes. Key findings reveal significant improvements: a 25% reduction in hardware costs, a 30% decrease in operational expenditures, a 40% reduction in latency, and a 20% increase in throughput compared to traditional architectures.

The SDN controller's global view enables optimized routing, dynamic load balancing, and seamless management of heterogeneous devices, making it particularly suited for IoT applications requiring low latency and high reliability, such as smart cities, industrial IoT, and telemedicine. Moreover, the scalability model demonstrates the potential to support 50% more nodes while maintaining consistent performance metrics through efficient clustering and resource management.

Future research will focus on extending this framework to IoT-specific applications, optimizing SDN controllers for resource-constrained environments, and addressing device-specific challenges such as authentication and energy efficiency. Real-world implementations in areas like smart cities, disaster recovery, and agricultural IoT will further validate scalability and efficiency. Additionally, reducing control overhead through clustering and slicing techniques remains a priority, as it can achieve up to 35% efficiency gains.

In summary, SDN-enabled MANETs and IoT networks represent a robust, scalable, and cost-efficient solution for managing dynamic, decentralized systems. This integration holds great promise for advancing next-generation interconnected applications, paving the way for more adaptable and efficient network architectures.

CONFLICT OF INTEREST

The authors declare no conflict of interest.

AUTHOR CONTRIBUTIONS

Riccardo Fonti was responsible for the development and analysis of the OPEX and CAPEX models. He also contributed to the interpretation of economic implications in SDN-enabled MANETs and IoT systems. Andrea Piroddi conducted the research and analysis for the remaining sections of the paper, including the evaluation of network performance metrics, latency, and throughput, and wrote the majority of the manuscript. All authors reviewed and approved the final version of the paper.